\documentclass[12pt]{article}
\usepackage{color}
\usepackage{subfigure}
\usepackage{epsfig}
\usepackage{amssymb,amsmath}

\begin{document}

\begin{titlepage}
\begin{flushright}
\today
\end{flushright}
\vfill
\begin{centering}

{\bf Compatibility of neutrino DIS data and global analyses of parton distribution functions}

\vspace{0.5cm}
Hannu Paukkunen\footnote{hannu.paukkunen@usc.es} and 
Carlos~A. Salgado\footnote{carlos.salgado@usc.es}

\vspace{1cm}
{Departamento de F\'\i sica de Part\'\i culas and IGFAE, Universidade de Santiago de 
Compostela, Galicia--Spain}

\vspace{1cm} 
{\bf Abstract} \\ 
\end{centering}

Neutrino$\backslash$antineutrino deep inelastic scattering (DIS) data provide
useful constrains for the flavor decomposition in global fits of parton
distribution functions (PDF). The smallness of the cross-sections requires
the use of nuclear targets in the experimental setup. Understanding the nuclear
corrections is, for this reason, of utmost importance for a precise determination
of the PDFs. Here,
we explore the nuclear effects in the
neutrino$\backslash$antineutrino-nucleon DIS by
comparing the NuTeV, CDHSW, and CHORUS cross-sections
to the predictions derived from the latest parton
distribution functions and their nuclear modifications. 
We obtain a good description of these data and find no apparent disagreement
between the nuclear effects in neutrino DIS and those
in charged lepton DIS. These results also indicate that further improvements in the knowledge of the nuclear PDFs could be obtained by a more extensive use of these sets of neutrino data.

\noindent

\vfill
\end{titlepage}

\setcounter{footnote}{0}

\section{Introduction}
\label{sec:intro}

The undisputed success of the recent global analyses of the free nucleon parton
distribution functions (PDFs) \cite{Nadolsky:2008zw,Martin:2009iq,Ball:2010de} have proven
the QCD factorization theorem as a well-working tool to analyse and interpret
the experimental cross-sections measured at high-energy colliders. These
analyses look for a universal set of PDFs, $f_{i}^A(x,Q^2)$, facilitating
the cross-sections computations schematically as
\begin{equation}
\sigma_{AB\rightarrow h+X} = \sum_{i,j} f_{i}^A(Q^2) \otimes f_{j}^B(Q^2) \otimes \hat{\sigma}_{ij\rightarrow h+X} , 
\label{eq:factorization}
\end{equation}
where $\hat{\sigma}_{ij\rightarrow h+X}$ are perturbatively computable coefficients.

The usual procedure to obtain a global fit of PDFs is well
established\footnote{A new approach with a different treatment of
the initial conditions, especially designed to propagate experimental
errors, has been recently proposed \cite{Ball:2010de}.}: a functional
form with $N$ parameters is assumed for the different
PDFs at a given initial scale $Q_0^2$. This initial condition is then evolved
by the Dokshitzer-Gribov-Lipatov-Altarelli-Parisi (DGLAP) equations \cite{DGLAP} and
the best set of parameters obtained in an iterative procedure involving a quality
criterion --- a particular definition of the goodness of the fit through
a generalized $\chi^2$. Several differences among the analyses, like the
choice of the initial conditions, the choice of the data sets, the actual
definition of the $\chi^2$ or the approximations in the computation of the
actual cross-sections, result in differences in the obtained PDFs.
In particular, the neutrino$\backslash$antineutrino DIS data is included only in part of
these analyses due to the difficulty in removing the nuclear effects
in a model-independent manner. These data are, however, of importance
for the flavor decomposition of the PDFs which, on the other hand, affects
the precision in the calculation of relevant cross-sections in present
accelerators, in particular the LHC.

In the situation, where the nucleons are part of a bound nucleus,
the factorization theorem could be theoretically more doubtful, as other processes
like multiple scattering could ruin its applicability. 
However, the corresponding global analyses
\cite{Eskola:2009uj,Schienbein:2009kk,Hirai:2007sx,deFlorian:2003qf}
show also an excellent description of charged-lepton deep inelastic scattering (DIS) and Drell-Yan (DY) dilepton production
data involving nuclear targets with the assumption of universal, process-independent, nuclear PDFs (nPDFs). This universality has also been checked in other processes, as single-inclusive hadron production in dAu collisions at RHIC, first introduced in \cite{Eskola:2008ca} and then also in \cite{Eskola:2009uj}. The access to harder and harder probes in nuclear collisions made the need of this type of analyses, and the corresponding checks of Eq. (\ref{eq:factorization}), also of utmost importance for the phenomenological interpretation of the imminent LHC heavy-ion program. 

The amount of experimental information used in global nPDF fits is much smaller than the corresponding one for the free proton case. The total number of data points is ${\cal O}(1000)$ scanning the range of $x\gtrsim 10^{-2}$ in the perturbative region for different nuclei. These data are normally given in terms of the ratio of nuclear over free proton cross-sections (with the appropriate normalization factor) so most of the global analyses study the corresponding ratios of PDFs $R_i^A(x,Q^2) \equiv f_i^{\rm bound}(x,Q^2)/f_i^{\rm free}(x,Q^2)$. In other words, the benchmarking role of the proton (or deuteron) cross-section in the experimental data is played in the global fits by a known set of free proton PDFs. 

On the other hand, the abundant neutrino DIS data available for nuclear targets, both iron and lead, could provide strong constrains to the global fits of nuclear PDFs. The absence of a corresponding proton benchmarking data in this case indicates, however, that the price to pay is a modification of the usual procedure to check the compatibility among the different data sets within a DGLAP analysis. Several attempts to study nuclear effects in neutrino DIS exist \cite{Hirai:2004ba,Eskola:2006ux,Kulagin:2007ju,Armesto:2007tg,Schienbein:2007fs}. It has been recently pointed out \cite{Schienbein:2007fs}, that the NuTeV neutrino-Iron DIS data \cite{Tzanov:2005kr}
displays a behaviour that hints to a possible disagreement between the nuclear modifications
$R_i^A(x,Q^2)$ in neutrino DIS, and those extracted from charged lepton DIS and DY process. If true,
this would seriously doubt the validity of QCD factorization for bound nucleons. This is not only a problem for the nuclear PDF analyses, but would also make these data not suitable to be included in free proton global fits, and relevant constrains for different flavors would be lost.

In this work we address this issue by studying at the same time the compatibility of neutrino DIS data on iron and lead targets from three experiments within the approach of factorizable universal PDFs. These PDFs are taken from the latest available sets CTEQ6.6 \cite{Nadolsky:2008zw} for protons and EPS09  \cite{Eskola:2009uj} for nuclei, in which these data have not been included. We use, in particular, data from three collaborations NuTeV \cite{Tzanov:2005kr}, CDHSW \cite{Berge:1989hr}, and
CHORUS \cite{Onengut:2005kv}. The error analyses performed for PDFs allow us to propagate the uncertainties from other sets of data and to check the compatibility both between data and theory and among different sets of experimental data.
In order to present the data in a more transparent manner, we provide as benchmark cross-sections for the nuclear case the ones computed theoretically using the known set of free proton PDFs. Our analysis finds a good description of neutrino DIS with nPDFs fitted to other data sets and do not find support for a breaking of the universality of the nuclear PDFs. We also comment on the compatibility between different data sets, in particular concerning the systematics in energy of the $x$-dependence in NuTeV data.

\section{Framework}
\label{sec:framework}

\subsection{Computation Framework}
\label{sec:Computation_Framework}

A general expression for the differental DIS cross-sections can be written down in terms of three nuclear structure functions
$F_i^{\nu A, \overline\nu A}$ as
\begin{eqnarray}
 \frac{d^2\sigma^{\nu A, \overline\nu A}}{dxdy} = & & \frac{G_F^2 M_W^4}{\left(Q^2+M_W^2\right)^2}
  \frac{M E_\nu}{\pi} \label{eq:GeneralCrossSection} \\
  & & \left[ xy^2 F_1^{\nu A, \overline\nu A} + \left( 1-y-\frac{xyM}{2E_\nu} \right) F_2^{\nu A, \overline\nu A} \pm
   xy\left(1-\frac{y}{2} \right) F_3^{\nu A, \overline\nu A} \right]. \nonumber
\end{eqnarray}
where $y=Q^2/(2xME_\nu)$, $G_F = 1.16637 \times 10^{−5} \, {\rm GeV}^{−2}$ is the Fermi constant,
$M = 0.938 \, {\rm GeV}$ denotes the nucleon mass, $M_W= 80.398 \, {\rm GeV}$ the mass of the $W$ boson, and $E_\nu$ to the neutrino energy. The `` + '' sign in front of the last term
refers to the neutrino induced process, whereas `` -- '' sign corresponds to the antineutrino scattering.

Each of the nuclear structure functions are composed of those of the bound protons and neutrons 
\begin{equation}
 F_i^A = \frac{Z}{A} F_i^{{\rm proton},A} + \frac{A-Z}{A} F_i^{{\rm neutron},A}.
\end{equation}
We calculate the structure functions in the next-to-leading order QCD-improved parton model, schematically
as a convolution between the perturbative Wilson coefficients $\omega$ and the PDFs $f_k^{{\rm proton},A}$
\begin{equation}
 F_i^{{\rm proton},A} = \sum_k \omega_{ik} \otimes f_k^{{\rm proton},A},
\end{equation}
where the $k$ runs over all parton flavors. The neutron structure functions
$F_i^{{\rm neutron, A}}$ are computed correspondingly by assuming the isospin symmetry,
i.e.
\begin{equation}
f_u^{{\rm neutron},A} = f_d^{{\rm proton},A}, \quad f_d^{{\rm neutron},A} = f_u^{{\rm proton},A}.
\end{equation}
The effects of the heavy quarks
are taken into account by the SACOT-prescription \cite{Aivazis:1993kh,Aivazis:1993pi,Kramer:2000hn,Kretzer:2003it,Tung:2006tb}.
This is also the scheme which was adopted by CTEQ6.6 \cite{Nadolsky:2008zw} analysis. This set will be used here to compute the benchmark cross-sections for the neutrino-nucleon cross-sections.

The nuclear modifications to the free proton PDFs $f_k^{{\rm proton}} = f_k^{{\rm CTEQ6.6}}$ are
taken from the recent EPS09-analysis\footnote{The EPS09-analysis was performed in the zero-mass scheme.
However, in the presently studied region of $x$, we have checked that the difference between general-mass and zero-mass
schemes in $R_k^A(x,Q^2)$ are small.} \cite{Eskola:2009uj}, which relates the nuclear PDFs to the
free proton ones by a multiplicative factor $R_k^A(x,Q^2)$ as
\begin{equation}
 f_k^{{\rm proton},A}(x,Q^2) = R_k^A(x,Q^2) \, f_k^{{\rm proton}}(x,Q^2).
\end{equation}

\subsection{Experimental Input}
\label{sec:ExpData}

\begin{figure}[!htb]
\center
\includegraphics[scale=0.7]{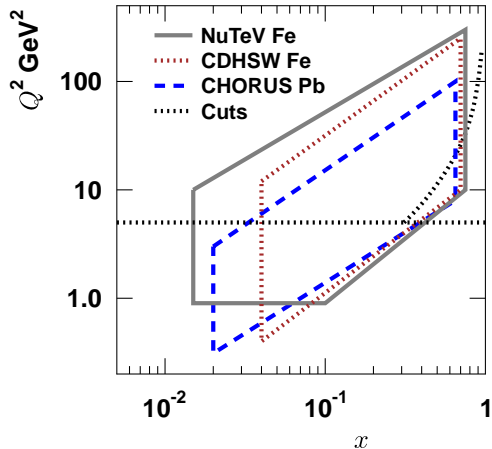}
\caption[]{\small The kinematical reach of the NuTeV, CDHSW, and CHORUS neutrino data. The
dotted black lines are the kinematical borderlines of Eq.~(\ref{eq:kincut}).}
\label{Fig:Data}
\end{figure}

The experimental data on which we base our analysis are differential charged-current
neutrino$\backslash$antineutrino-nucleus cross-sections\footnote{For a review of neutrino experiments,
the reader can consult e.g. Ref.~\cite{Conrad:1997ne}.} ${d^2\sigma^{\nu A, \overline\nu A}}/{dxdy}$
measured at several values of neutrino beam energies $E_\nu$, Bjorken-$x$, and virtuality $Q^2$.
The NuTeV collaboration \cite{Tzanov:2005kr} provides 2618 and CDHSW \cite{Berge:1989hr} 1533 data points
from Iron targets, whereas the CHORUS collaboration \cite{Onengut:2005kv} gives 1214 data points from Lead target
measurements. The kinematical reach of these data are illustrated by the polygons of Figure~\ref{Fig:Data}. 
The CTEQ analysis which is used here as reference for the proton PDFs imposes cuts in experimental data which will be also taken into account here.
For consistency with this analysis and to avoid
getting contaminated from higher-twist effects we restrict the virtuality $Q^2$ and the final state invariant
mass $W^2 = M^2 + Q^2(1-x)/x$ by the kinematical boundaries
\begin{equation}
Q_{\rm cut}^2 > 5 \, {\rm GeV}^2, \qquad W_{\rm cut}^2  > 12 \, {\rm GeV}^2. \label{eq:kincut} 
\end{equation}
These cuts are indicated by black dottet lines in Figure~\ref{Fig:Data}. After applying
these restritions, 2041 NuTeV data points, 700 CDHSW, and 768 CHORUS data points remain.

\subsection{Target Mass and Radiative Corrections}

The two main types of additional corrections factors that should
be applied before a meaningful comparison to the 
experimental data are the target mass correction,
and the one originating from the electroweak radiation.

The target mass corrections are important at large $x$ and low $Q^2$. In this work,
we apply the collinear factorization prescription
from Ref.~\cite{Accardi:2008ne}, which basically amounts to changing the Bjorken-$x$
variable to the Nachtmann variable, defined as $\xi \equiv 2x/(1+ \sqrt{1+4x^2M^2/Q^2})$.
To be precise, we make the substitution
\begin{equation}
 \int_x^1 \frac{dz}{z} \omega_{ik} \left( z \right) f_k^A \left( \frac{x}{z} \right)
 \rightarrow
 \int_x^1 \frac{dz}{z} \omega_{ik} \left( z \right) f_k^A \left( \frac{\xi}{z} \right)
 \label{eq:targetmass}
\end{equation}
for the subprocesses involving light partons.

In contrast to the target mass corrections, the radiative corrections are rather a small-$x$
issue and more spread out in $Q^2$. To account for the radiative
corrections, we use the parton flavor dependent factors $\Delta_k^{\rm radiative}$ from
Ref.~\cite{Arbuzov:2004zr} correcting the leading order terms, i.e. we compute
\begin{equation}
 F_i^{A} = \sum_k \left[\omega^{\rm LO}_{ik} \left(1 + \Delta_k^{\rm radiative}\right) + \omega^{\rm NLO}_{ik}\right] \otimes f_k^A.
\end{equation}

Both corrections modify the $Q^2$-dependence of the calculated cross-sections. 
In the next sections, we will study the $Q^2$-dependence of the data vs. theory ratios and 
and by examining them, one can judge
whether obvious additional ingredients outside pQCD remain ---  only those regions
of phase space where such behaviour is not evident, can be reliably used to make
conclusions about the nuclear PDFs.

\subsection{Probed PDF Components}

To get a rough understanding of the behaviour of the neutrino cross-sections
and their sensitivity to the different PDF-components, it is sufficient to
consider the leading order expressions for the structure functions. For simplicity,
we suppress the Cabibbo-Kobayashi-Maskawa matrix elements and assume that one is
at large enough $Q^2$ to forget about the heavy quark mass effects. Then, for
the neutrino scattering
\begin{eqnarray}
  F_2^{\nu A} = 2x F_1^{\nu A} & = & 2x \left( d^A + s^A + b^A + \overline u^A + \overline c^A \right) \\
  F_3^{\nu A} & = & \,\,\,\, 2 \left( d^A + s^A + b^A - \overline u^A - \overline c^A \right),
\end{eqnarray}
where e.g.
\begin{eqnarray}
d^A & = & \frac{Z}{A} d^{\rm proton,A} + \frac{A-Z}{A} d^{\rm neutron,A} \\
    & = & \frac{Z}{A} d^{\rm proton,A} + \frac{A-Z}{A} u^{\rm proton,A}, \nonumber
\end{eqnarray}
assuming the isospin symmetry. For the antineutrino scattering
\begin{eqnarray}
  F_2^{\overline\nu A} = 2x F_1^{\overline\nu A} & = & 2x \left( u^A + c^A + \overline d^A + \overline s^A + \overline b^A \right) \\
  F_3^{\overline\nu A} & = & \,\,\,\, 2 \left( u^A + c^A - \overline d^A - \overline s^A - \overline b^A \right).
\end{eqnarray}
By inserting in Eq.~(\ref{eq:GeneralCrossSection}), one finds that the cross-sections
depend on the following partonic combinations:
\begin{eqnarray}
d^2\sigma^{\nu A} & \propto & \left(d^A+s^A+b^A\right) + \left(1-y\right)^2 \left(\overline u^A + \overline c^A \right) \label{eq:Par_nu} \\
d^2\sigma^{\overline\nu A} & \propto & \left(\overline d^A +\overline s^A + \overline b^A\right) + \left(1-y\right)^2 \left(u^A + c^A \right). \label{eq:Par_anu}
\end{eqnarray}
Because of the suppressing $(1-y)^2$ factor, the antineutrino cross-section carries more
sensitivity to the sea quarks towards larger $Q^2$, whereas the neutrino cross-sections always
get a significant contribution from the valence quarks.

\subsection{The Method Of Comparison}
\label{sec:Analysis_Principle}

In what follows, we will present the comparison to the data in terms of
the following data vs. theory ratios:
\begin{equation}
 R^{\rm CTEQ6.6} \equiv \frac{\sigma^{\nu,\overline\nu}\left({\rm Experimental} \right)}{\sigma^{\nu,\overline\nu}\left({\rm CTEQ6.6} \right)},
\end{equation}
where CTEQ6.6 denotes the set of PDF used in the calculation. The PDF
uncertainties are considered by calculating the upper and the lower
cross-section uncertainties $\Delta \sigma^\pm$ by a formula
\begin{eqnarray}
(\Delta \sigma^\pm)^2 & = & \sum_k \left[ 
\begin{array}{c}
 \max  \\
 \min
\end{array}
\left\{ \sigma(S^+_k)-\sigma(S^0), \sigma(S^-_k)-\sigma(S^0),0 \right\} \right]^2, \label{eq:error_best}
\end{eqnarray}
where $\sigma(S^0)$ is the value of the cross-section computed by the CTEQ6.6 central set, and
$\sigma(S_k^\pm)$ are the ones calculated by the $k$th uncertainty sets. We will compare such ``pseudodata''
to the expected nuclear effects using the nuclear modification factors $R_k^A(x,Q^2)$ from EPS09-analysis
\begin{equation}
 R^{\rm CTEQ6.6 \times EPS09} \equiv \frac{\sigma^{\nu,\overline\nu}\left({\rm CTEQ6.6 \times EPS09} \right)}{\sigma^{\nu,\overline\nu}\left({\rm CTEQ6.6} \right)}, \label{eq:nuclear_mod_npdfs}
\end{equation}
which allow us to make a conclusion about the consistency of the nuclear
effects between neutrino scattering and the charged lepton scattering.
We note that the EPS09 includes a possibility for a similar uncertainty analysis
as in Eq.~(\ref{eq:error_best}). The size of the uncertainties of this origin will be addressed
in section~\ref{sec:Other_Sets_of_nPDFs}.

\section{Results and Discsussion}
\label{sec:Results}

\subsection{The $\chi^2$-values}
\label{sec:Chi2}

The standard way of measuring the agreement between the experimental data
and the theory, is to calculate the $\chi^2$ defined as
\begin{equation}
 \chi^2 \equiv \sum_{i \in {\rm data \, points}} \left( \frac{T_i-D_i}{\sigma_i} \right)^2,
\end{equation}
where $T_i$ is the theoretical value, and $D_i$ is the corresponding experimental 
value for the $i$th data point with estimated uncertainty $\sigma_i$ \footnote{The experimental data used here have been taken from the database \cite{HEPDATA}. Statistical and systematic errors have been added in quadrature. Additional overall normalization errors have been ignored.}. In this work
we add the statistical and systematic uncertainties in quadrature when computing
$\chi^2$. Usually, if the $\chi^2/N$ ($N$ being the number of data points) is not much
larger than one, the theoretical calculations are considered as being statistically consistent
with the data.

\begin{table}
\begin{center}
{\footnotesize
\begin{tabular}{ccc}
 RAD + TM & CTEQ6.6 &  CTEQ6.6$\times$EPS09 \\
\hline
\hline
 NuTeV	& 1.51  & 1.05 \\
 CHORUS & 1.15  & 0.79 \\
 CDHSW  & 1.10  & 0.71 \\
 \\
RAD + No TM & CTEQ6.6 &  CTEQ6.6$\times$EPS09 \\
\hline
\hline
 NuTeV	& 1.29  & 1.02 \\
 CHORUS & 1.03  & 0.85 \\
 CDHSW  & 0.88  & 0.66 \\
 \\
No RAD + No TM & CTEQ6.6 &  CTEQ6.6$\times$EPS09 \\
\hline
\hline
 NuTeV	& 1.35  & 1.08 \\
 CHORUS & 1.23  & 1.09 \\
 CDHSW  & 0.96  & 0.86  \\
\end{tabular}
}
\caption[]{\small The $\chi^2/N$-values computed using CTEQ6.6 with and without nuclear
modification from EPS09. The numbers are given for calculations including radiative
and target mass correction (RAD + TM), with radiative but without target
mass correction (RAD + No TM), and without radiative nor target mass corrections (No RAD + No TM).}
\label{Table:chi2values}
\end{center}
\end{table}

In Table~\ref{Table:chi2values}, we give the $\chi^2/N$-values, computed
with those data points that meet our kinematical constraints.
We have calculated the values in three different ways using CTEQ6.6 PDFs with and
without nuclear modifications from EPS09: With the radiative and the target mass corrections,
with the radiative but without target the mass correction, and finally without the radiative nor
the target mass corrections. We make the following observations:
\begin{itemize}
 \item 
Whatever way we make the calculation, the one without
nuclear corrections from EPS09 gives substantially larger $\chi^2$.
 \item
For CHORUS and CDHSW data, the $\chi^2$s get consistently smaller as the radiative and
target mass are applied.
 \item
For NuTeV data, the $\chi^2$ remains practically unchanged whether radiative or
target mass corrections are applied or not.
\end{itemize}

Whereas the good $\chi^2/N$-values indicate that the theory calculations
are 
in good agreement
with the data, it may seem puzzling that while
$\chi^2$s for CHORUS and CDHSW gain improvement when applying the
radiative and the target mass corrections, the $\chi^2_{\rm NuTeV}$ stays
essentially constant.

The fact that the $\chi^2_{\rm CHORUS}$ and $\chi^2_{\rm CDHSW}$ get clearly
smaller after addition of the radiative and target mass corrections indicates, that
there are a substantial changes in the computed cross-sections. Therefore,
the unchanging $\chi^2_{\rm NuTeV}$ implies that there is something peculiar
in the NuTeV data sample.

\begin{figure}[!htb]
\center
\includegraphics[scale=0.7]{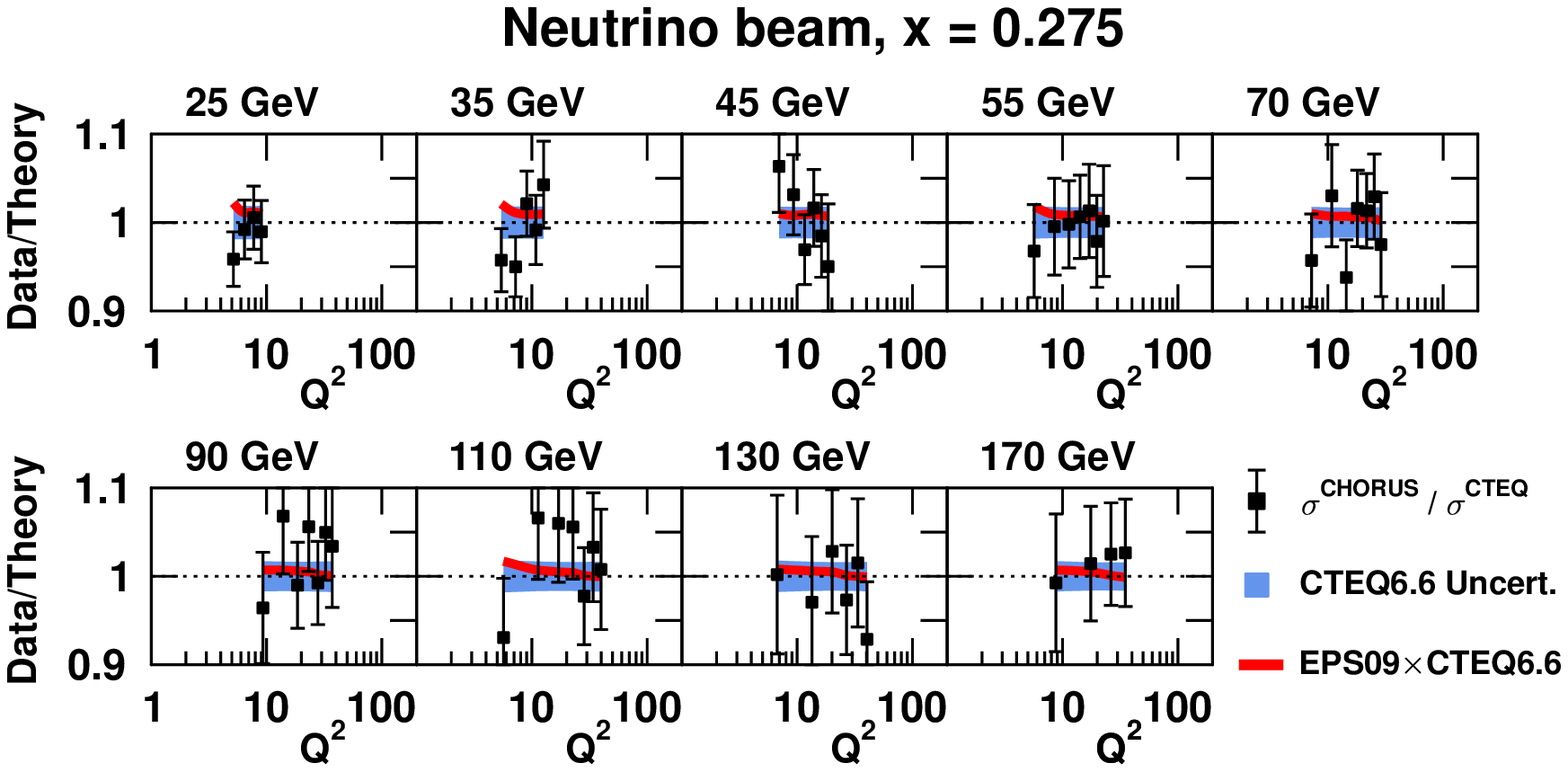}
\includegraphics[scale=0.7]{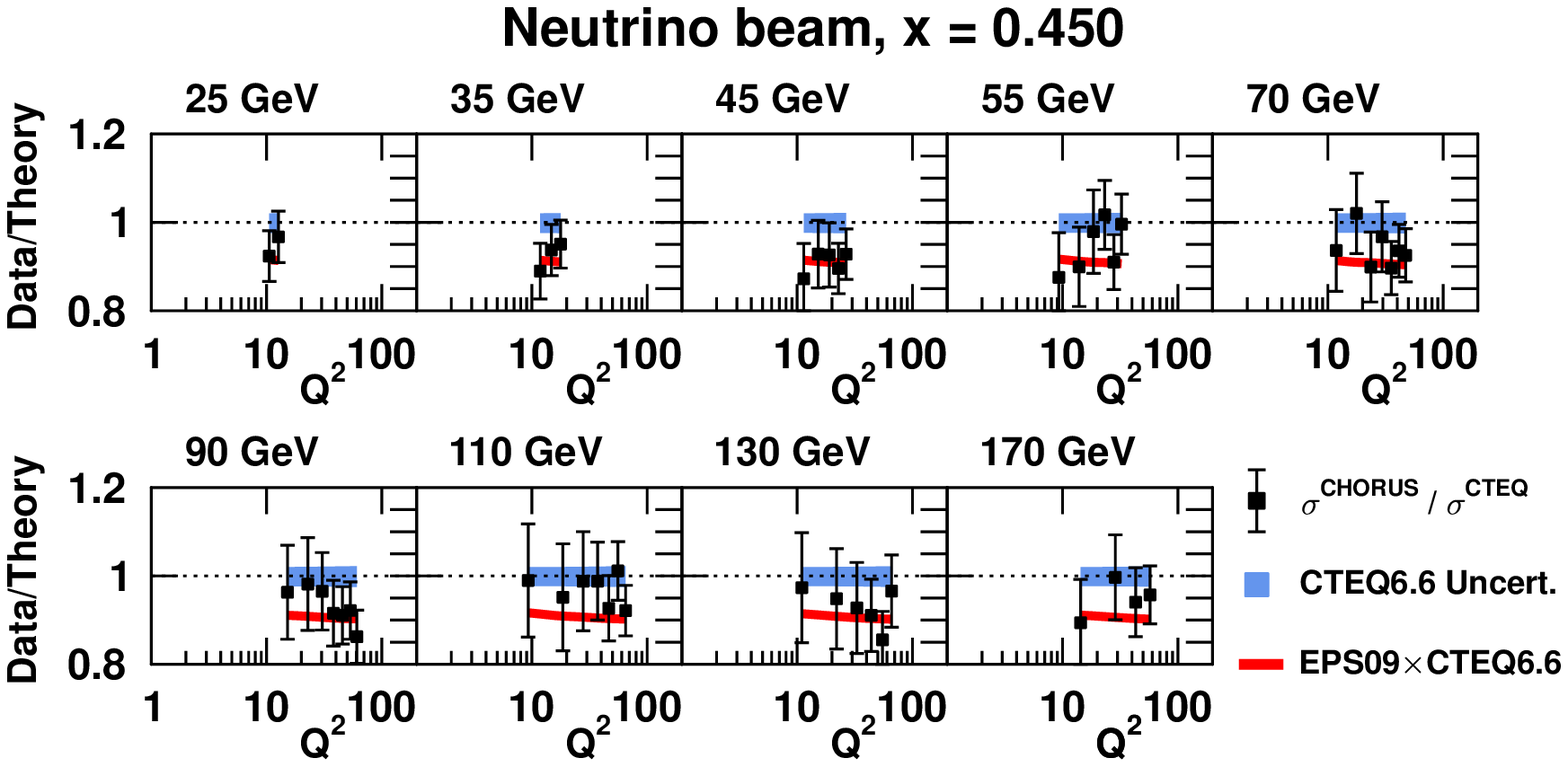}
\caption[]{\small Examples of Data/Theory ratios for the CHORUS neutrino data.
The black squares with error bars correspond to the ratios $R^{\rm CTEQ6.6}$,
whereas the red lines are the $R^{\rm CTEQ6.6\times EPS09}$ predictions. The
relative CTEQ6.6 uncertainty is shown by blue bands.}
\label{Fig:Ratio1}
\end{figure}

We note that the target mass correction, Eq. (\ref{eq:targetmass}), was not applied
in the EPS09-analysis. Repeating the analysis including such effect
would slightly change the shape of the valence quark nuclear modifications
at large-$x$. Moreover, the CTEQ6.6 uncertainties also grow towards large $x$.
Both effects, if taken into account, would modify the $\chi^2$ evaluation of the data in this region of phase space. Therefore, the changes in $\chi^2$ when
applying the target mass corrections for the CTEQ6.6$\times$EPS09 column
of Table \ref{Table:chi2values} should be taken only as indicative. Note, however, how the $\chi^2$ gets clearly
worse when adding the target mass correction if the nuclear PDFs are not used.

\subsection{Data vs. Theory Ratios}
\label{sec:Data_Theory_Ratios}

\begin{figure}[!htb]
\center
\includegraphics[scale=0.75]{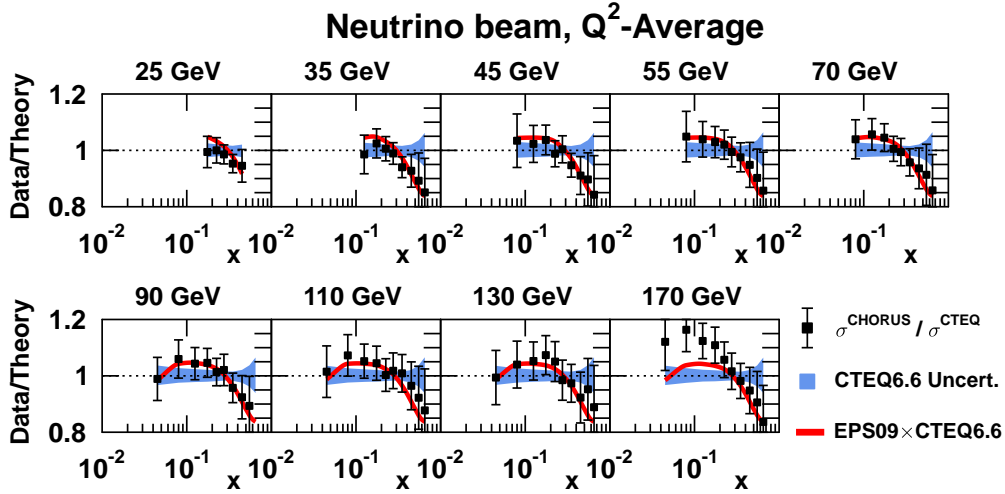}
\caption[]{\small The $Q^2$-averaged CHORUS neutrino data. The black squares
with errorbars show the $R_{\rm Average}^{\rm CTEQ6.6}$ ratios, while the
red lines correspond to  $R_{\rm Average}^{\rm CTEQ6.6 \times EPS09}$. The blue bands
denote the average relative uncertainty in the cross-sections $d^2\sigma^\nu({\rm CTEQ6.6})$.}
\label{Fig:Q_nu_average_CHORUS}
\end{figure}
\begin{figure}[!htb]
\center
\includegraphics[scale=0.75]{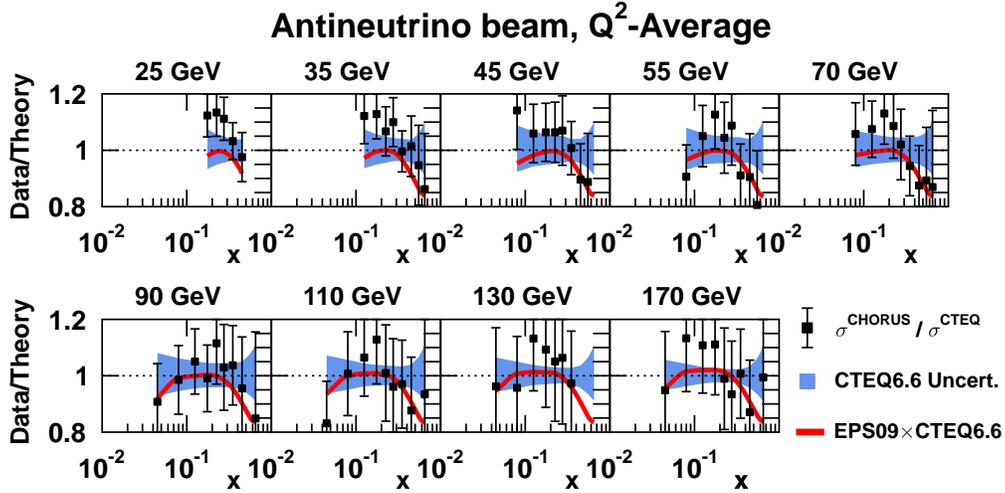}
\caption[]{\small As Figure~\ref{Fig:Q_nu_average_CHORUS}, but for CHORUS antineutrino data.}
\label{Fig:Q_anu_average_CHORUS}
\end{figure}
\begin{figure}[!htb]
\center
\includegraphics[scale=0.75]{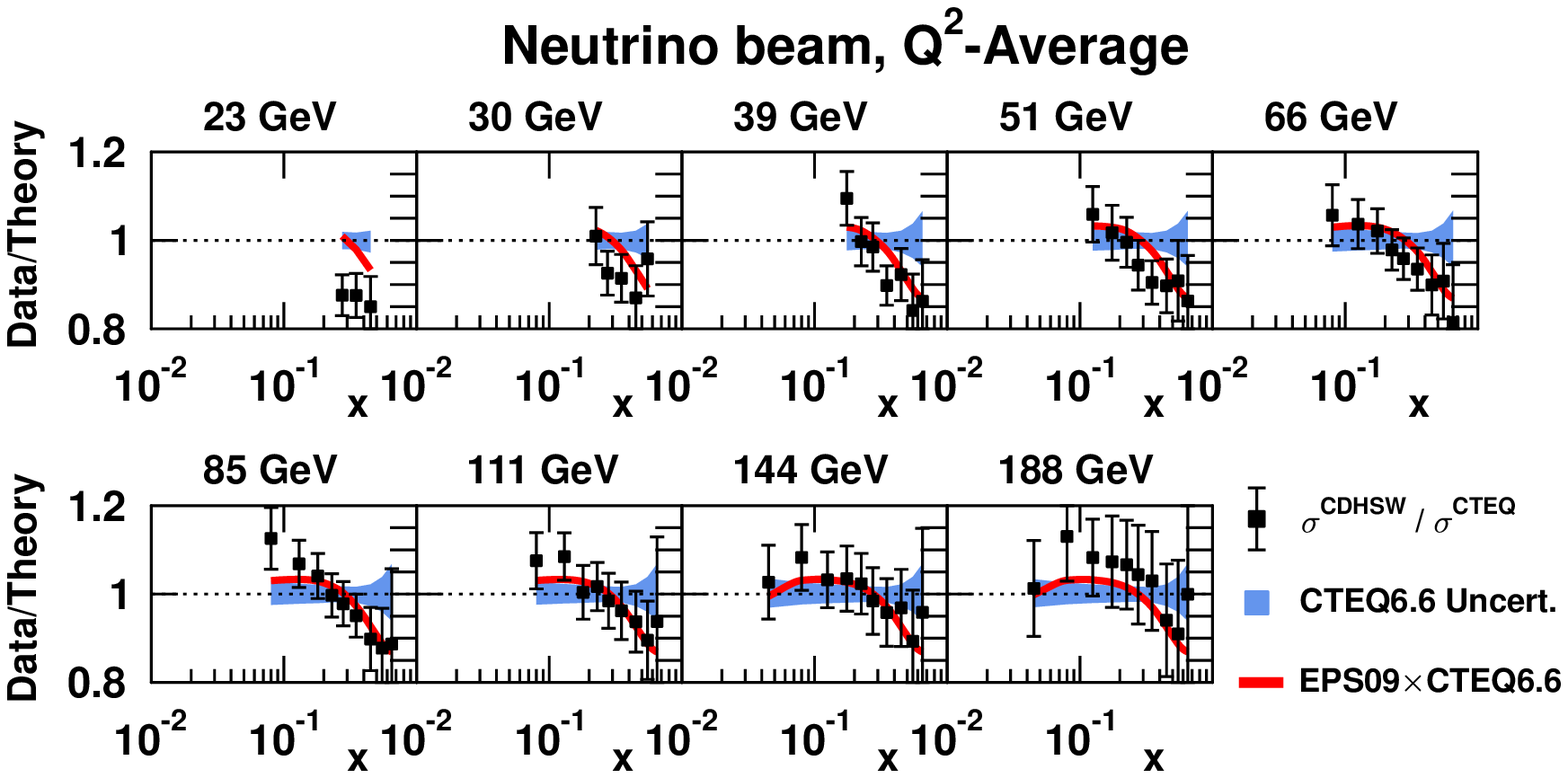}
\caption[]{\small As Figure~\ref{Fig:Q_nu_average_CHORUS}, but for CDHSW neutrino data.}
\label{Fig:Q_nu_average_CDHSW}
\end{figure}
\begin{figure}[!htb]
\center
\includegraphics[scale=0.75]{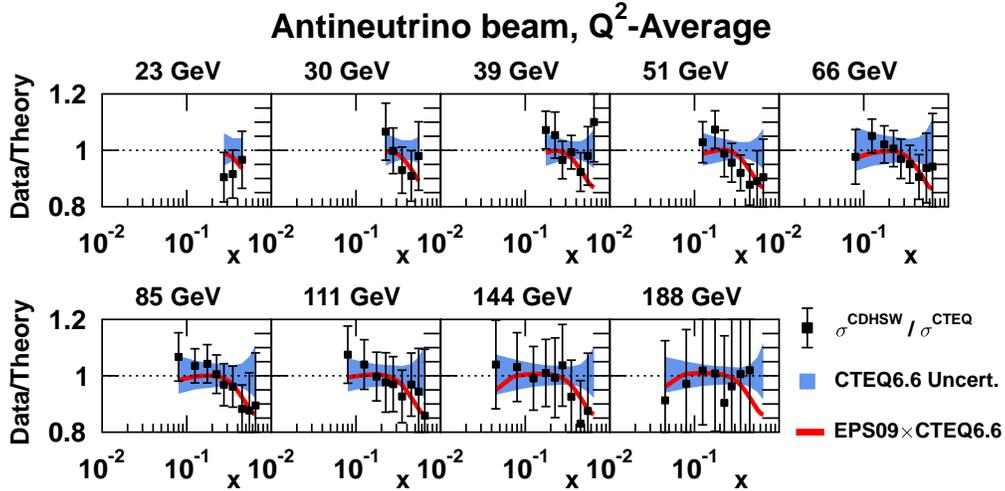}
\caption[]{\small As Figure~\ref{Fig:Q_nu_average_CHORUS}, but for CDHSW antineutrino data.}
\label{Fig:Q_anu_average_CDHSW}
\end{figure}
\begin{figure}[!htb]
\center
\includegraphics[scale=0.7]{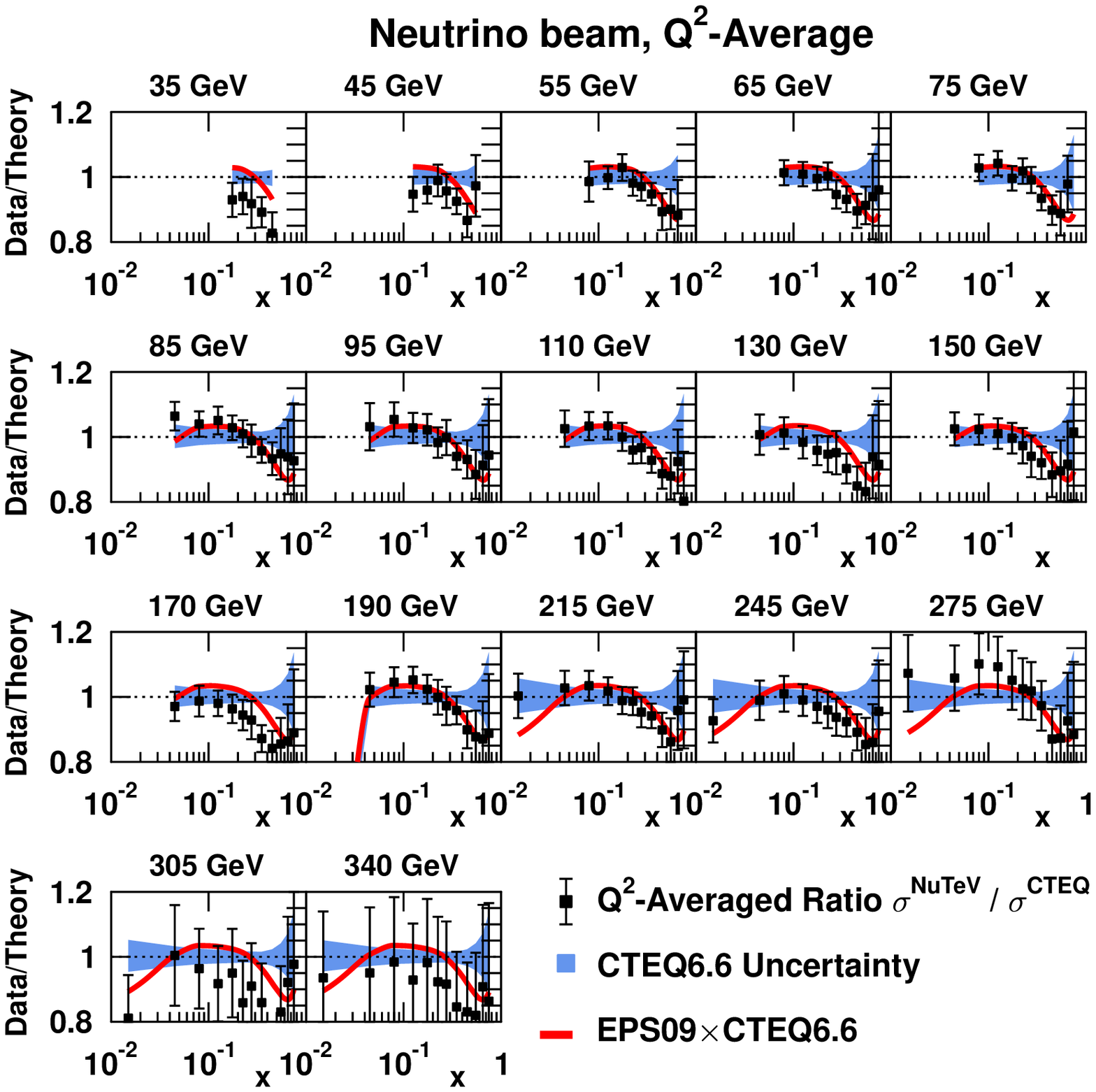}
\caption[]{\small As Figure~\ref{Fig:Q_nu_average_CHORUS}, but for NuTeV neutrino data.}
\label{Fig:Q_nu_average_NuTeV}
\end{figure}
\begin{figure}[!htb]
\center
\includegraphics[scale=0.7]{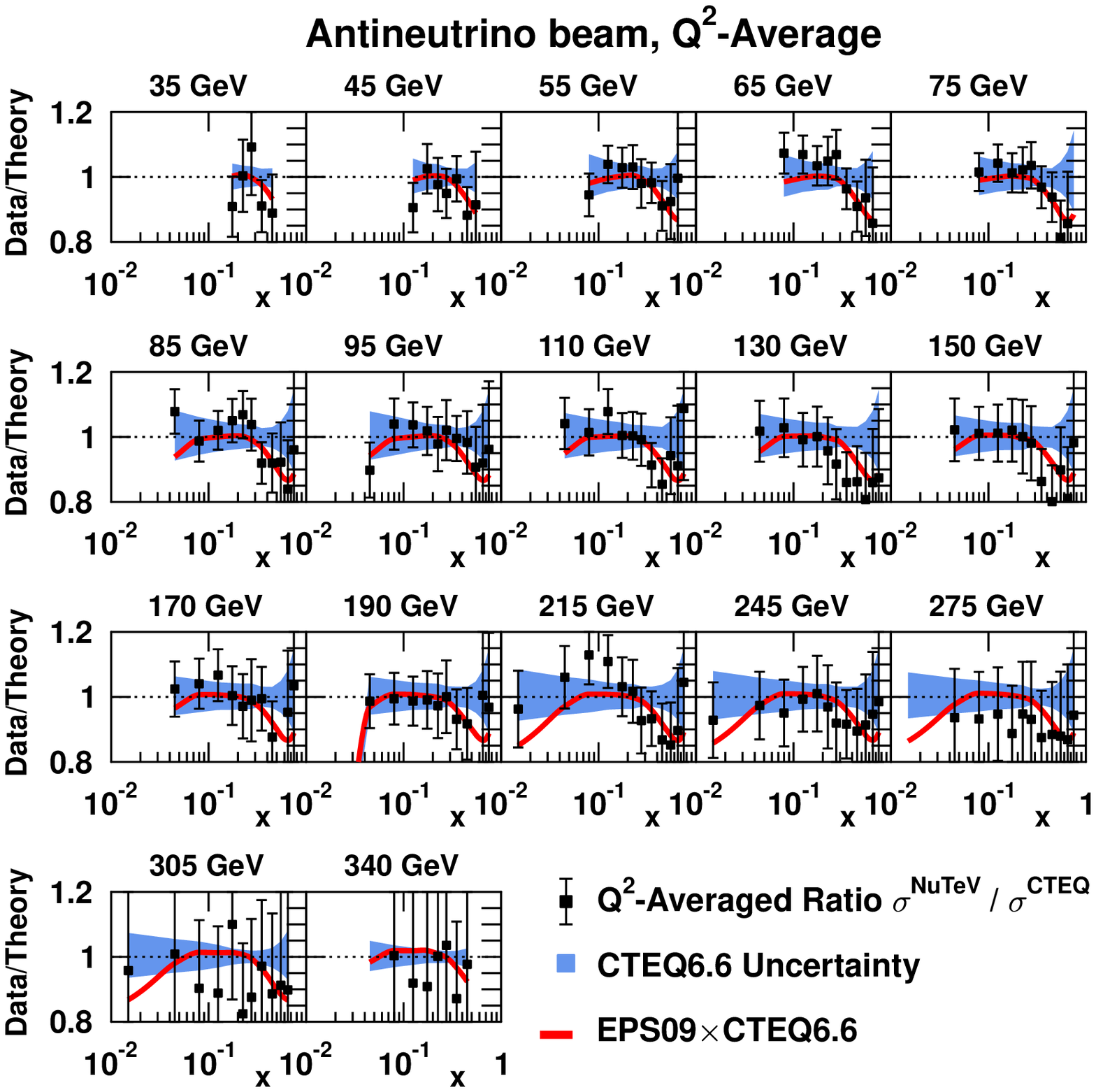}
\caption[]{\small As Figure~\ref{Fig:Q_nu_average_CHORUS}, but for NuTeV antineutrino data.}
\label{Fig:Q_anu_average_NuTeV}
\end{figure}

In this section we take a closer look to the nuclear effects
in neutrino-nucleus DIS by plotting results for ratios
$R^{\rm CTEQ6.6}$ and $R^{\rm CTEQ6.6 \times EPS09}$. To begin with, the
Figure~\ref{Fig:Ratio1} shows two representative examples of the
energy and $Q^2$-dependence of the data vs. theory ratios. Evidently,
as one cannot observe a large systematic $Q^2$-dependence in these panels, it is
reasonable to look a $Q^2$-averaged versions of these figures in order to 
better see the big picture --- the $x$-shape of the nuclear modifications.

We will form the $Q^2$-averages of the panels by the prescription
\begin{equation}
 R^{\rm CTEQ6.6}_{\rm Average} \equiv
 \left( \sum^N_{i\in {\rm fixed} \, x} \frac{R_i^{\rm CTEQ6.6}}{\sigma_i} \right) \left( \sum^N_{i\in {\rm fixed} \, x} \frac{1}{\sigma_i} \right)^{-1}
 \pm
 N \times \left( \sum^N_{i\in {\rm fixed} \, x} \frac{1}{\sigma_i} \right)^{-1},
\end{equation}
where the sum runs over all data points in same $x$- and energy-bin, but different $Q^2$.
We note that the way we compute the average and, especially its uncertainty, is not
meant to be statistically ``the correct way'', but it is rather ment to give an idea
about the size of the uncertainties. We stress, however, that the statistical analysis (the calculation of the $\chi^2$-values in the previous sections) was performed with the original (non-averaged) data. The purely theoretical calculations are simple averages,
for example
\begin{equation}
 R^{\rm CTEQ6.6 \times EPS09}_{\rm Average} \equiv \frac{1}{N} \sum^N_{i\in {\rm fixed} \, x} R_i^{\rm CTEQ6.6 \times EPS09}. \label{eq:av_nuclear_mod_npdfs}
\end{equation}
The results of this exercise are presented
in Figures~\ref{Fig:Q_nu_average_CHORUS}-\ref{Fig:Q_anu_average_NuTeV}.
We wish to draw the reader's attention to the following points:

\begin{itemize}
\item
The CTEQ6.6$\times$EPS09-prediction and the CHORUS and CDHSW neutrino data are in a perfect agreement.
The data do not display an evident dependence of the neutrino energy. This is also true
for the CDHSW antineutrino data.
\item
The CHORUS antineutrino data seems to somewhat overshoot the CTEQ6.6$\times$EPS09-prediction
around $x \sim 0.2$ staying, however, clearly in the uncertainty limits.
\item
Whereas some panels corresponding to the NuTeV neurino data are very well in line with
the CTEQ6.6$\times$EPS09 prediction, there is an evident variation from a beam energy
to another. For example $E_\nu = 170 \, {\rm GeV}$ and $E_\nu = 190 \, {\rm GeV}$
are in the brink of disagreement. 
\item
Similar evident energy-dependece is not observable in the NuTeV antineutrino panels.
Within the uncertainties the data are well reproduced.
\end{itemize}

The clear energy-dependent panel-to-another variation in the NuTeV neutrino data/theory ratios is
evidently something that cannot be compensated by only re-fitting the PDFs --- the CTEQ6.6$\times$EPS09
predictions vary very slowly as a function of the incident neutrino
energy. On the contrary, using this data set as a source of PDFs in a fit would inevitably result in a
compromise between data in different energy bins displaying mutual tension.
No signs of this kind of evident controversy are visible in the NuTeV antineutrino data nor in
the CHORUS or CDHSW data.

This observation also explains why the $\chi^2_{\rm NuTeV}$ is almost inert
to the target mass and radiative corrections. Indeed, by simply ignoring the
$E_\nu=35, 45, 130, 150, 170, 245 \, {\rm GeV}$ panels when computing  $\chi^2_{\rm NuTeV}$,
would cause the $\chi^2_{\rm NuTeV}/N$ to reduce from 1.10 to 0.94 when applying the corrections.

\subsection{Other Sets of nPDFs}
\label{sec:Other_Sets_of_nPDFs}

\begin{figure}[!htb]
\center
\includegraphics[scale=0.5]{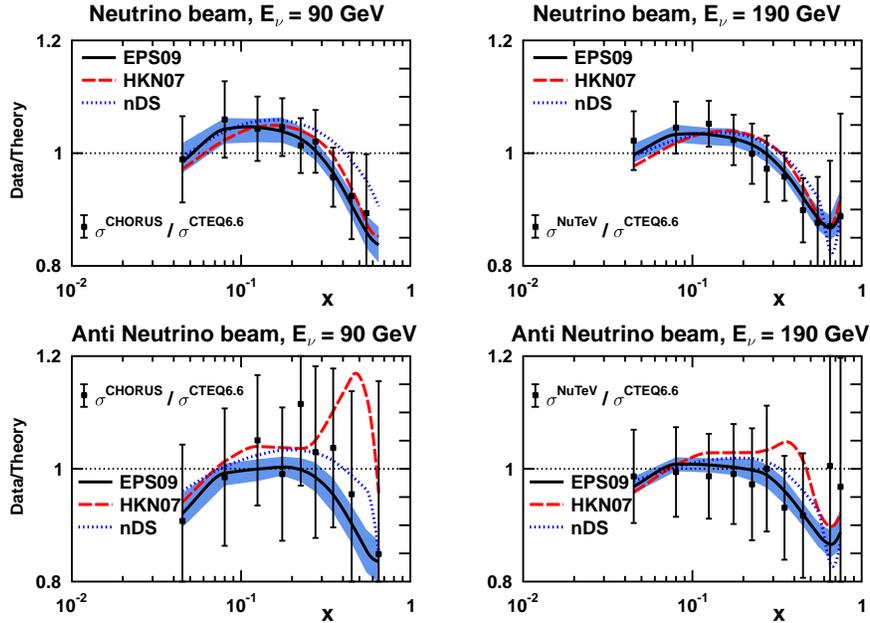}
\caption[]{\small Two panels of $Q^2$-averaged CHORUS and NuTeV data with predictions from three
sets of nPDFs, EPS09 (black line with blue errorband), HKN07 (red dashed line), and nDS
(blue dotted line).}
\label{Fig:Compare_nPDFs}
\end{figure}

So far we have considered the nuclear effects in PDFs utilizing solely the
latest available analysis, EPS09, observing a good average agreement with
the experimental data. To address whether the same conclusion would have been drawn
using different sets of nPDFs we show, in Figure~\ref{Fig:Compare_nPDFs}, examples
of the $Q^2$-averaged ratios from the two other available NLO sets nDS and HKN07, and
compare them to the uncertainty band of EPS09.

The upper panels illustrate the situation with the neutrino cross-sections.
At large $x$, the dominant contribution to the cross-sections comes from the
valence quarks and --- due to good amount of valence quark constrains from
charged lepton DIS there --- the different sets of nPDFs give predictions
relatively close to each other. Only the nDS prediction for Lead display
a clear deviation from others. Towards small $x$, the predictions
from different sets of nPDFs are very close to each other. 

In the case of antineutrino cross-sections, illustrated in the lower panels,
there is an enhanced contribution from the sea quarks,
for which the nuclear modifications at large $x$ are not very well constrained.
This uncertainty is reflected as clearly larger scatter between the curves.
At small $x$, all sets of nPDFs are again in a good agreement.

In summary, it is mainly in the large-$x$ sector where the differences between sets
of nPDFs become apparent: While the nDS description admits some improvements especially
regarding the $A$-dependence in this region, the HKN07, on the other hand clearly misses
the structure of antineutrino cross-sections at large-$x$. This is due to the radically different
behavior of antiquarks and gluons when compared with the two other sets in the corresponding
relevant region of phase space. These findings point also to the adequacy of neutrino data in
providing further constrains for nPDFs.

\section{Conclusion}
\label{Conclusion}

In this article we have presented a study of the nuclear effects in the
(anti)neutrino-nucleus deeply inelastic scattering. As a conclusion, we
find that from the analysed data sets only one --- the NuTeV neutrino data ---
shows signs of not being in a complete agreement with the present-day PDFs. 
This discrepancy, we argue, is however something that cannot be completely
cured by simply re-fitting the PDFs. Rather, it is due to an unexplained,
neutrino energy-dependent fluctuations in the data --- in order to make this point more explicit, we have presented $Q^2$-averaged data where the energy systematics of the $x$-dependence is more transparent. Therefore we assert
that this data set cannot be taken as a discriminating factor when making
conclusions about universality of the nuclear PDFs.
The forthcoming data from NOMAD experiment will eventually
facilitate a further clarification about the conflict between the different
experiments.

Apart from this finding, the present analysis do not give any reason to
belive that the nuclear effects in neutrino-nucleus DIS would
be different from those extracted from charged lepton DIS and DY
production of dileptons --- the factorization seems to work well.

Having said that, there is a lot of room for improvement
and future work. For example, the EPS09-analysis leaves
no freedom to the strange quark nuclear modification, but it is something that is fixed by
hand to other sea quarks. As the antineutrino cross-sections carry
a non-neglibible contribution from the strange quarks, including these data to
a global fit of nPDFs would offer a chance to free this assumption made in EPS09.
This could improve the slight undershooting of the CHORUS antineutrino data.
Also, whereas the Iron is almost an isoscalar nucleus ($Z=26$, $N=30$),
the Lead is clearly non-isocalar ($Z=82$, $N=126$). This could offer a possibility to
separately constrain the nuclear modifications for the up and down quarks which
has not been possible before. 

\section*{Acknowledgments}
We thank N\'estor Armesto for useful discussions. 
This work is supported by Ministerio de Ciencia e Innovaci\'on of Spain under project FPA2009-06867-E; by Xunta de Galicia (Conseller\'{\i}a de Educaci\'on and Conseller\'\i a de Innovaci\'on e Industria -- Programa Incite); by the Spanish Consolider-Ingenio 2010 Programme CPAN (CSD2007-00042); and by by the European Commission grant
PERG02-GA-2007-224770. CAS is a Ram\'on y Cajal researcher.

\end{document}